\documentclass[twocolumn,showpacs,preprintnumbers,amsmath,amssymb]{revtex4}
\usepackage{graphicx}
\usepackage{dcolumn}
\usepackage{bm}
\newcommand{\Eq}[1]{Eq.~({\protect\ref{#1}})}
\newcommand{\Ref}[1]{Ref.\protect\cite{#1}}
\newcommand{\Fig}[1]{Fig.~\protect\ref{#1}}

\newcommand{\Tate}{\rule{0cm}{1.1em}}
\newlength{\Tatescale}
\newcommand{\Hs}{\hspace*{1em}}
\newcommand{\Bs}{\hspace*{-0.5em}}

\newlength{\figwidth}
\newcommand{\BAr}[1]{\overline{#1}}

\begin{document}
\title{A practical  approach to the  problem of the  missing imaginary
part\\  of   the  handbag  diagram  in   the  confined  Bethe-Salpeter
framework}
\author{Noriyoshi ISHII}
\email{ishii@rarfaxp.riken.go.jp}
\affiliation{Radiation Lab.,
RIKEN (The Institute of Physical and Chemical Research),\\
2-1 Hirosawa, Wako, Saitama 351-0198, Japan
}

\date{\today}

\begin{abstract}
In the  confined Bethe-Salpeter (BS) methods, the  quark propagator is
an entire function.  Hence, the  imaginary part of the handbag diagram
disappears, leading to a  problem of the vanishing parton distribution
function (PDF).  In contrast, the direct calculation of the light-cone
(LC) momentum  distribution does give  a non-vanishing result  even in
the  confined BS framework.   We consider  their precise  relation and
difference, and propose  to use the latter as  a practical approach to
this  problem.  Our  formalism  is  general enough  to  be applied  to
various effective models.
\end{abstract}
% 11.10.St, Bound and unstable states; Bethe-Salpeter equations
% 13.60.Hb, Total and inclusive cross sections (including deep-inelastic processes)
% 12.39.Ki, Relativistic quark model
% 12.40.Yx, Hadron mass models and calculations
%
\maketitle
%%%%%%%%%%%%%%%%%%%%%%%%%%%%%%%%%%%%%%%%%%%%%%%%%%%%%%%%%%%%%%%%%%%%%%

The  parton  distribution  function   (PDF)  is  defined  through  the
factorization  procedure  in the  perturbative  QCD  (pQCD)  as a  low
energy-scale quantity which cannot  be calculated within the framework
of pQCD  itself \cite{parton}.
It provides  us with a unique  tool to extract the  information on the
quark-gluon structure of hadrons directly based on experiments.
In  order  to calculate  PDF  theoretically,  the  information of  the
hadronic  wave  function  plays  the  essential role.   Hence,  it  is
necessary  to resort  to various  low-energy  nonperturbative methods,
such  as low  energy effective  models, the  light-front quantization,
lattice QCD, etc.

The  Bethe-Salpeter (BS)  method is  one such  method, which  makes it
possible to evaluate a number  of hadronic low-energy observables in a
relativistically  covariant manner  \cite{williams,tandy,lorenz}.  The
BS equation has been applied to mesons which consist of mainly a quark
and an antiquark \cite{williams,tandy,lorenz,kugo,knaito}.
The BS description  of baryons as the relativistic  three quark system
has been  developed in the  last decade, utilizing the  Faddeev method
for        the         quantum        three        body        problem
\cite{ishii1,Hanhart:1995tc,asami,lorenz}.     The    Faddeev   method
transforms the  three-quark BS equation into  the relativistic Faddeev
equation, which justifies the use  of the quark-diquark BS equation as
its approximation\cite{ishii1,buck,mineo,oettel,laith,lorenz}.

In the  BS method, the  effect of the  color confinement can  be taken
into account through the entire function nature of the nonperturbative
quark propagator  \cite{williams,tandy,lorenz,roberts}.  Thereby, many
of the shortcomings  can be improved such as  the unphysical threshold
for decay  into colored  particles, description of  higher resonances,
etc.
However,  there is  actually a  serious drawback  in  calculating PDF.
Note that,  to calculate PDF in  the BS method,  one usually considers
the  handbag diagram first  \cite{roberts,toki,kulagin,kusaka}.  Then,
PDF is realized as the Bjorken limit of its imaginary part. In most of
the  low energy  effective  models, this  procedure  has been  adopted
providing reasonable  answers.  In the  confined BS framework,  due to
the entire  function nature of  the propagator, the imaginary  part of
the handbag  diagram disappears, leading  to a serious problem  of the
vanishing PDF.

A straightforward solution of this problem may be found by considering
the intermediate ``hadronic loops'', which can provide imaginary parts
due to the  color-singlet nature of hadrons.  If we  adopt this as the
solution,  we have  to  consider the  full  forward Compton  amplitude
instead of the handbag diagram.
Remember that, in  order to calculate the full  Compton amplitude, one
has    to   adopt    the   recently    developed    ``gauge   method''
\cite{kvinikhidze,ishii,ariola},  which determines  the unique  set of
Feynman  diagrams   associated  with  the  particular   choice  of  BS
interaction  kernel. (Other  choice of  Feynman diagrams  lead  to the
violation of Ward identity.)
It follows  that, in  order to  obtain the hadronic  loop, one  has to
solve a highly complicated BS equation.  This is of course formidable.
Its  numerical  solution is  practically  impossible  to be  obtained.
Moreover,  the construction  of the  BS equation  itself  is extremely
non-trivial  \cite{ishii.98,bender.96}.  The  transparent  relation to
the parton picture will be lost as well.

In this way,  one has to figure out another  practical approach to PDF
in the confined BS framework.
Actually, there is  another method for PDF by  resorting to the direct
calculation of the light-cone (LC) momentum distribution.
Since  it is  essentially  the  wave function  squared  with the  plus
component of  the LC momentum of one  of the partons fixed,  it is not
expected to vanish unless the wave function itself vanishes.
This procedure  has been mainly adopted in  the light-front framework,
where the LC momentum distribution is naturally expressed as the equal
light-front time correlator.
On the other hand, in the effective models (including BS method), this
procedure  has  been  scarcely   adopted  except  for  a  few  authors
\cite{mineo,thomas1991}.

The latter method is conceptually  sound in the sense that the Bjorken
limit should not be considered  within the framework of the low energy
effective theories.
The product of the currents should  be factorized at the level of pQCD
into the  product of the  high energy scale quantity  (the coefficient
functions) and low scale  quantity (the composite operators, i.e., the
LC momentum distributions).  The BS method should be used to calculate
the matrix element of the low energy quantities.
%%%
Then, the structure function $F_2(x,Q^2)$ is expressed for large $Q^2$
as   $F_2(x,Q^2)   \simeq   x\left(\frac1{9}  u(x,Q^2)   +   \frac1{9}
\BAr{u}(x,Q^2) +  \frac{4}{9} d(x,Q^2) +  \frac{4}{9} \BAr{d}(x,Q^2) +
\cdots\right)$  with  the  $Q^2$-evoluted  LC  momentum  distributions
$u(x,Q^2)$, $\BAr{u}(x,Q^2)$, $d(x,Q^2)$, $\BAr{d}(x,Q^2)$, etc.
%%%%
Fortunately, in  most of the  cases so far,  these two methods  do not
result  in a  serious  inconsistency.  Especially,  if the  propagator
satisfies  the scaling  property \cite{ariola},  both methods  lead to
exactly the  same result. This seems  to be the reason  why the latter
has  been  scarcely adopted  in  the  effective  models. However,  the
results of  these two methods do not  agree at all in  the confined BS
framework.
%%%%%%%%%%%%%%%%%%%%%%%%%%%%%%%%%%%%%%%%%%%%%%%%%%%%%%%%%%%%%%%%%%%%%%%%%%%%%%%

We begin by applying the inverse Mellin transformation analytically to
the matrix elements of the twist-two operators. The aim of doing so is
twofold.
The  first aim  is to  ensure that  the direct  calculation of  the LC
momentum distribution gives the equivalent  result with the one in the
moment  space.  As  is  well-known, pQCD  establishes the  equivalence
between  PDF  in  the  moment  space representation  and  PDF  in  the
$x$-space representation, where the equivalence is provided  by the Mellin
transformation.
The  second aim  is to  introduce the  quantities such  as  the ``{\it
forward   Compton-like  amplitude}''   and  the   ``{\it  handbag-like
diagram}'',  which will be  used to  consider the  precise differences
between the two methods for  PDF. We will argue their physical meaning
later.

We  consider  the  spin  averaged  matrix  element  of  the  twist-two
operators as follows:
\begin{equation}
	A_n
\equiv
	\frac1{2}
	\sum_{s=\pm}
	\frac{
		\langle N(p,s)|
			\BAr{\psi} \gamma^+ (i\partial^+)^{n-1}\psi
		|N(p,s)\rangle
	}{
		(p^+)^n
	},
\label{moment}
\end{equation}
where   $\psi$  and   $\BAr{\psi}$  denote   the  quark   fields,  and
$|N(p,s)\rangle$  denotes  the  state  vector  for  the  nucleon  with
momentum $p$  and helicity $s$.
The   light-cone   variables    are   defined   as   $p^{\pm}   \equiv
\frac1{\sqrt{2}}(p^0  \pm p^3)$, etc.
We   adopt   the   covariant  normalization   $\displaystyle   \langle
N(p,s)|N(p',s')\rangle = 2 E_N(p)  (2\pi)^3 \delta^{(3)}(\vec p - \vec
p')\delta_{ss'}$ with  $\displaystyle E_N(p)\equiv \sqrt{m_N^2  + \vec
p^2}$.
%%%%%%%%%%%%%%%%%%%%%%%%%%%%%%%%%%%%%%%%%%%%%%%%%%%%%%%%%%%%%%%%%%%%%%%%%%%%%%%
% curry curry curry
%%%%%%%%%%%%%%%%%%%%%%%%%%%%%%%%%%%%%%%%%%%%%%%%%%%%%%%%%%%%%%%%%%%%%%%%%%%%%%%
\begin{figure}[t]
\includegraphics[width=0.6\figwidth]{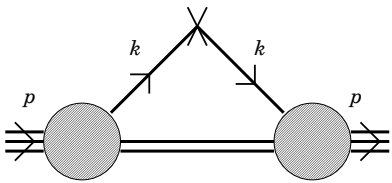}
\caption{The  diagram for  the matrix  element of  twist-two operators
$A_n$. The blob  followed by the triple-line denotes  the BS amplitude
for the  nucleon, and  the single-line denotes  the propagator  of the
quark.   The cross  symbol  ``$\times$'' represents  the insertion  of
$\gamma^+(k^+)^{n-1}/(p^+)^n$.}
\label{figure.an}
\end{figure}
The  actual  calculation of  the  matrix  elements  of these  bilinear
operators in the BS framework  amounts roughly to the diagram depicted
in \Fig{figure.an}.  The cross  ``$\times$'' indicates the point where
we insert the following vertex as
\begin{equation}
	\frac{\gamma^+(k^+)^{n-1}}{(p^+)^n}.
\label{cross}
\end{equation}
(For  more  details, see  \cite{kvinikhidze,ishii},  where the  matrix
element  of the  conserved current  is  discussed based  on the  gauge
method  \cite{kvinikhidze,ishii,ariola}, which  may  work as  explicit
examples. \Ref{asami} is also helpful.)

\begin{figure}[t]
\includegraphics[width=0.4\figwidth]{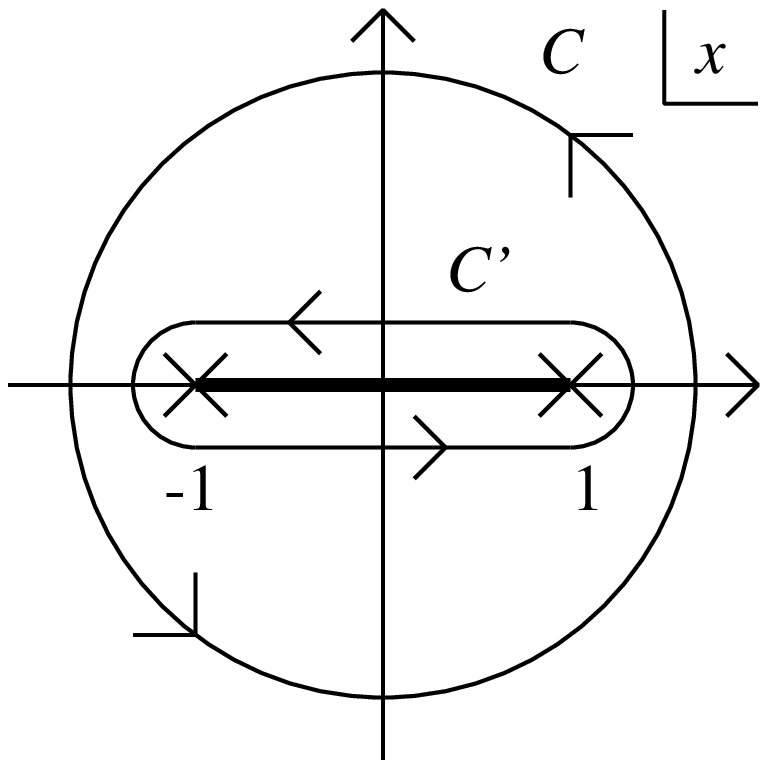}
\caption{The contour of the integral in \Eq{contour.integral}.}
\label{fig.contour}
\end{figure}

%%%
We apply analytically the  inverse Mellin transformation to $A_n$.
For this purpose, we introduce  an analytic function $T(x)$ for $x \in
\mathbb{C}$ in the following way:
\begin{equation}
	T(x)
\equiv
	\frac1{2\pi i}
	\sum_{n=1}^{\infty}
	\frac{A_n}{x^n}.
\label{compton-like}
\end{equation}
Since  $T(x)$  is  the  analogue  of  the  forward  Compton  amplitude
$T_{\mu\nu}(x,Q^2)$  in pQCD,  we refer  to $T(x)$  as  ``{\it forward
Compton-like amplitude}''.  We will argue its physical meaning later.
To proceed, we have to make  two assumptions on $T(x)$.  (1) The power
series  in  \Eq{compton-like}  converges,  if  $|x|  >  1$.   (2)  The
singularity structure of $T(x)$ is the cut along the segment $[-1,1]$.
Due to the assumption (1), in  the region $|x| > 1$, $T(x)$ defines an
analytic function,  which can be analytically continued  to the inside
as much as possible.
The  most important  property  of the  forward Compton-like  amplitude
$T(x)$ is the following identity:
\begin{equation}
	A_n
=
	\int_C dx\; x^{n-1} T(x),
\label{contour.integral}
\end{equation}
where the contour  $C$ is a sufficiently large  circle centered at the
origin as  depicted in \Fig{fig.contour}.  This  identity follows from
the residue theorem.
Due to the  assumption (2), we can deform the contour  $C$ to the thin
contour $C'$  as depicted in  \Fig{fig.contour}.  We are thus  lead to
the following identity:
\begin{equation}
	A_n
=
	\int_{-1}^{1} dx\; x^{n-1} H(x),
\label{desired}
\end{equation}
where $H(x)$ is the discontinuity of $T(x)$ along the cut, i.e.,
\begin{eqnarray}
	H(x)
&\equiv&
	\mbox{disc}\left(\Tate T(x) \right)
\label{inverse.mellin.transform}
\\\nonumber
&=&
	T(x-i\epsilon) - T(x+i\epsilon).
\end{eqnarray}
We define the quark and  the anti-quark distribution functions for $x$
($0 \le x \le 1$) as
\begin{equation}
	q(x)
\equiv
	H(x),
\Hs
	\BAr{q}(x)
\equiv
	-H(-x),
\label{pdf}
\end{equation}
respectively \cite{jaffe}. Now, \Eq{desired} reads
\begin{equation}
	A_n
=
	\int_0^1 dx x^{n-1}
	\left(\Tate
		q(x)
	+
		(-1)^n \BAr{q}(x)
	\right).
\label{sum.rule}
\end{equation}
From the relation to the  moments, this identity shows that $q(x)$ and
$\BAr{q}(x)$  really work  as PDF.   We note  that $H(x)$  is directly
expressed in  the canonical  operator representation in  the following
way \footnote{In  some papers, $T$-product is adopted  for the bilocal
operator.   As  is discussed  in  \Ref{jaffe},  for  the twist-two  LC
momentum distribution, $T$-product gives the identical result with the
ordinary product.}:
\begin{equation}
\renewcommand{\arraystretch}{1.5}
	H(x)
=
	\Bs
	\begin{array}[t]{l}\displaystyle
		\frac1{2}
		\sum_{s=\pm}
		\int_{-\infty}^{\infty} {dz^- \over 2\pi}
		e^{ixp^+ z^-}
	\\\displaystyle
	\times
		\left\langle N(p,s)\left|
			\BAr{\psi}(0) \gamma^+ \psi(z^-)
		\right| N(p,s)\right\rangle,
	\end{array}
\label{canonical.hx}
\end{equation}
where     $\psi(z^-)$     is      a     shorthand     notation     for
$\psi(z^+=0,z^-,z_{\perp}=0)$.  Indeed, by inserting \Eq{canonical.hx}
into  \Eq{desired}, we  obtain  \Eq{moment}.
%%%%%%%%%%%%%%%%%%%%%%%%%%%%%%%%%%%%%%%%%%%%%%%%%%%%%%%%%%%%%%%%%%%%%%

\begin{figure}[t]
\includegraphics[width=0.6\figwidth]{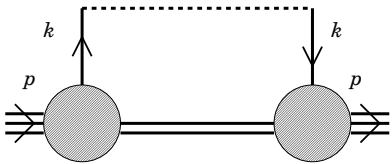}
\caption{The  handbag-like diagram, which  corresponds to  the forward
Compton-like   amplitude   $T(x)$.    The   dashed   line   represents
$\displaystyle  \frac1{2\pi  i}\frac{\gamma^+}{x  p^+  -  k^+}$.   The
meanings of the  blob followed by the triple-line  and the single-line
are the same as in \Fig{figure.an}.}
\label{fig.compton-like}
\end{figure}

To see  explicitly that $q(x)$ and $\BAr{q}(x)$  in \Eq{pdf} calculate
the   LC  momentum   distribution,  we   complete  the   summation  in
\Eq{compton-like}   analytically.    Since   the   cross   symbol   in
\Fig{figure.an} represents  the vertex $\gamma^+ (k^+)^{n-1}/(p^+)^n$,
the summation  in \Eq{compton-like} reduces to the  summation of these
vertices as
\begin{eqnarray}
	V(x)
&=&
	\sum_{n=1}^{\infty}
	\frac1{2\pi i}
	\frac{\gamma^+ (k^+)^{n-1}}{x^n(p^+)^n}
\\\nonumber
&=&
	\frac1{2\pi i}
	\frac{\gamma^+}{x p^+}
	\sum_{n=0}^{\infty}
	\left(\frac{k^+}{xp^+}\right)^n
\\\nonumber
&=&
	\frac1{2\pi i}
	\frac{\gamma^+}{x p^+ - k^+}.
\label{bjorken.limit}
\end{eqnarray}
The diagram, which corresponds to the direct calculation of $T(x)$, is
depicted in  \Fig{fig.compton-like}, where the  dotted line represents
$V(x)$. Since this diagram is  the analogue of the handbag diagram, we
refer to it as ``{\it handbag-like diagram}''.
In order to  calculate $H(x)$, all we have to do  is to replace $V(x)$
by the following quantity:
\begin{eqnarray}
\lefteqn{
	V(x - i\epsilon) - V(x + i\epsilon)
}
\label{delta}
\\\nonumber
&=&
	\frac{\gamma^+}{2\pi i}
	\left(
		\frac1{x p^+ - k^+ - i\epsilon}
	-
		\frac1{x p^+ - k^+ + i\epsilon}
	\right)
\\\nonumber
&=&
	\gamma^+ \delta(x p^+ - k^+),
\end{eqnarray}
where  we  used  the  identity:  $\delta(x) =  \frac1{2\pi  i}  \left(
\frac1{x - i\epsilon} - \frac1{x + i\epsilon} \right).$
\begin{figure}[t]
\includegraphics[width=0.6\figwidth]{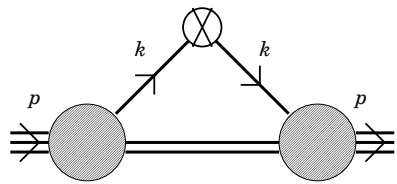}
\caption{The   diagram   which    represents   $H(x)$.    The   symbol
``$\otimes$''  represents the  point, where  $\gamma^+ \delta(x  p^+ -
k^+)$  is  inserted.  The  meanings   of  the  blob  followed  by  the
triple-line and the single-line are the same as in \Fig{figure.an}.}
\label{fig.delta}
\end{figure}
Since \Eq{delta}  counts the ``number''  of each parton  which carries
the LC momentum $xp^+$,  $q(x)$ and $\BAr{q}(x)$ in \Eq{pdf} calculate
directly the  LC momentum distribution.  By construction,  it is clear
that the quantity  which is equivalent to the  twist-two moments under
the  Mellin transformation  is the  PDF  obtained as  the LC  momentum
distribution rather than the PDF obtained from the handbag diagram.

We consider the  physical meaning of the handbag-like  diagram and the
precise  differences between  the two  methods.  We  may think  of the
handbag-like  diagram as  the ``{\it  proper Bjorken  limit}''  of the
handbag diagram.  Indeed, the  LC momentum distribution is obtained as
the  imaginary  part of  this  ``proper  Bjorken  limit'' of  the  handbag
diagram,  while, in  the original  method, the  imaginary part  of the
handbag diagram in the straightforward Bjorken limit is considered.
(We  emphasize  again that  these  two  methods  would agree,  if  the
propagator  of the  struck quark  could satisfy  the  scaling property
\cite{ariola}.)  We  can find the  precise difference between  the two
methods  in  the  difference  between  these  two  diagrams.   In  the
handbag-like  diagram, the  free  propagator is  used  for the  struck
quark, while,  in the handbag diagram,  the nonperturbative propagator
is  used.    Intuitively,  since  the  extremely   large  momentum  is
transfered to the  struck quark, there is no reason  to believe in the
effective   nonperturbative  propagator,   which  is   constructed  to
reproduce the low energy properties.  Furthermore, since the origin of
the  propagator  of  the  struck  quark  in  our  formulation  is  the
derivatives  in  the  twist-two   operators,  it  should  not  be  the
non-trivial one.
%% \footnote{Once  the   gluons  are  involved,   the  situation  becomes
%% subtle. The use of $A^+ = 0$ gauge may solve the problem. At any rate,
%% there  does not  seem to  be  so far  the effective  model, which  can
%% explicitly take into account the effect of the gluons.}.

Several  comments  are in  order.
%%%%%%%%%%
First, the nontrivial propagator should  be used except for the struck
quark.  In \Ref{roberts}, the first  attempt to calculate PDF for pion
in the confined  BS framework is performed, where  the free propagator
is used  not only  for the  struck parton but  also for  the spectator
parton.   Their calculation is  different from  our formalism  in this
way.
%%%%%%%%%%
Second,  by  construction,  our  LC  momentum  distribution  naturally
satisfies the  various sum rules,  i.e., the fermion number  sum rule,
the momentum  sum rule, and  so on.  (See \Eq{sum.rule}.   Its moments
calculate the matrix elements of various conserved currents.)
%%%%%%%%%%
The third comment  is concerning the calculability of  the LC momentum
distribution.  Usually, to evaluate  the LC momentum distribution, one
insert $\gamma^+\delta(x p^+ - k^+)$ as the vertex in the diagram like
\Fig{figure.an}.   In some  cases,  it  is not  easy  to perform  this
procedure  straightforwardly.   (For  instance, when  the  integration
momentum becomes complex due to  the Wick rotation, the meaning of the
delta function  with the complex  variable is highly  nontrivial.)  In
such  cases, the technique  of the  inverse Mellin  transformation can
extend   the   calculability   of   the   LC   momentum   distribution
\cite{mineo-dron}.   As is  often  the case,  the  calculation in  the
moment   representation  is   much  easier   than  the   one   in  the
$x$-representation.
%%%%%%%%%%
The last comment is concerning  the gluon. Once the gluon is attempted
to be explicitly taken into  account, the derivative in \Eq{moment} is
replaced  by  the covariant  derivative.   At  the canonical  operator
level,  the inverse Mellin  transformation leads  to \Eq{canonical.hx}
where the  path-ordered integral  $P\exp i\int_C A_{\mu}  dx^{\mu}$ is
inserted.  Here,  $C$ is the  light-like straight path  connecting $0$
and  $(z^{+}=0,z^{-},z_{\perp}=0)$.   At  the  BS  level,  it  becomes
nontrivial to perform analytically the Mellin transformation.  We note
however that, it is not the handbag-like diagram with its struck quark
propagator replaced  by the non-perturbative  one ---it would  be much
more  complicated.  In  this sense,  the straightforward  solution may
look attractive, since it makes  us to avoid this difficulty. However,
as  we  have  argued,  its  practical  implementation  is  formidable.
Another solution may be provided by $A^+ = 0$ gauge.

To summarize,  we have considered  the problem of  vanishing imaginary
part of the handbag diagram in the confined BS framework. We have seen
that a practical approach is provided by the direct calculation of the
LC  momentum   distribution.   In   order  to  consider   the  precise
differences  between  the  two   methods  for  PDF,  we  have  applied
analytically  the  inverse Mellin  transformation  to the  expectation
value  of the  twist-two operators.   We have  introduced  the forward
Compton-like  amplitude and  the handbag-like  diagram, which  are the
analogues of  the forward Compton  amplitude and the  handbag diagram,
respectively. In our formalism, the handbag-like diagram serves as the
``proper  Bjorken limit'' of  the handbag  diagram, and  its imaginary
part  calculates the  LC  momentum distribution,  i.e.,  the PDF.   We
finally remark that  our formalism is general enough  to be applied to
various  effective  models.    In  realistic  calculations,  we  often
encounter with the nontrivial  quark propagators, which cannot satisfy
the  scaling property.   Even  if such  a  nontrivial propagators  are
involved, the evaluation of PDF can be done with our method.

\begin{center}{\bf Acknowledgment}\end{center}
The  author  thanks  R.~Alkofer,  W.~Bentz  and  K.~Kusaka  for  their
fruitful discussions.  A part of this work was done during his stay in
Germany under the support by a grant of BMBF.

%%%%%%%%%%%%%%%%%%%%%%%%%%%%%%%%%%%%%%%%%%%%%%%%%%%%%%%%%%%%%%%%%%%%%%

\end{document}